\documentclass[%
  aps,
  pra,
  showkeys,
  twocolumn,
  superscriptaddress
]{revtex4-2}

\PassOptionsToPackage{%
  colorlinks=true,%
  linkcolor=red,%
  citecolor=blue,%
  urlcolor=blue%
}{hyperref}

\usepackage[utf8]{inputenc}
\usepackage[T1]{fontenc}
\usepackage{lmodern}

\usepackage{amsmath,amssymb,amsfonts}
\usepackage{mathrsfs}
\usepackage{braket}
\usepackage{bm}

\usepackage{array}
\usepackage{booktabs}
\usepackage{dcolumn}

\usepackage{graphicx,color}
\usepackage{float}

\usepackage{pstricks}
\usepackage{tikz}

\usepackage{orcidlink}
\usepackage{comment}

\usepackage{hyperref}


\begin{document}

\title{Controlled quantum secure remote sensing}
\author{Muhammad Talha Rahim \orcidlink{0000-0001-9243-417X}} \email{mura68827@hbku.edu.qa}
\affiliation{Qatar Center for Quantum Computing, College of Science and Engineering, Hamad Bin Khalifa University, Doha, Qatar}
\author{Saif Al-Kuwari \orcidlink{0000-0002-4402-7710}} \email{smalkuwari@hbku.edu.qa}
\affiliation{Qatar Center for Quantum Computing, College of Science and Engineering, Hamad Bin Khalifa University, Doha, Qatar}
\author{M.I. Hussain\orcidlink{0000-0002-6231-7746}}
\email{mahussain@hbku.edu.qa}
\affiliation{Qatar Center for Quantum Computing, College of Science and Engineering, Hamad Bin Khalifa University, Doha, Qatar}
\author{Asad Ali\orcidlink{0000-0001-9243-417X}} \email{asal68826@hbku.edu.qa}
\affiliation{Qatar Center for Quantum Computing, College of Science and Engineering, Hamad Bin Khalifa University, Doha, Qatar}

\date{\today}
\def\be{\begin{equation}}
\def\ee{\end{equation}}
\def\bea{\begin{eqnarray}}
\def\eea{\end{eqnarray}}
\def\f{\frac}
\def\n{\nonumber}
\def\l{\label}
\def\p{\phi}
\def\o{\over}
\def\R{\rho}
\def\pa{\partial}
\def\om{\omega}
\def\na{\nabla}
\def\P{\Phi}

\begin{abstract}
Quantum resources enable secure quantum sensing (SQS) of remote systems, offering significant advantages in precision and security. However, decoherence in the quantum communication channel and during the evolution of quantum states can erode these advantages. In this work, we first propose a general $N-$particle scheme that achieves Heisenberg-limited (HL) scaling for single-parameter estimation in the presence of an ideal quantum communication channel and encoding scenario. For non-ideal dynamics, we introduce a modified protocol incorporating local quantum optimal control (QOC) operations to address noise in SQS under generalized Pauli dephasing and parallel dephasing noise. We analyze two distinct scenarios: a noiseless communication channel with noisy evolution, and a noisy communication channel with noisy evolution. For the noisy channel, we model the link between the communicating parties as a depolarizing channel. The protocol leverages QOC operations to actively mitigate noise, enhancing the achievable quantum Fisher information (QFI) and the classical Fisher information (CFI) based on the chosen measurement strategy.
\end{abstract}


\maketitle


\section{Introduction}\label{sec:introduction}

Quantum metrology protocols offer the potential to achieve Heisenberg-limited (HL) precision measurements, surpassing the standard quantum limit (SQL) that constrains classical systems \cite{GLM:04:Science, GLM:06:PRL, GLM:11:Nature, TA:14:JPA, DM:14:PRL}. By harnessing key quantum phenomena, such as superposition and entanglement, these protocols significantly enhance the sensitivity of the measurement. Beyond improving precision, these quantum properties also play a critical role in quantum communication by providing inherent security, and therefore are also leveraged in communication protocols such as quantum key distribution \cite{SBC:09:RMP, XMZ:20:RMP} and quantum secure direct communication \cite{LDW:FPC:07, PSL:23:ADI} to achieve information-theoretic security.

To realize secure quantum sensing (SQS) of remote systems, hybrid approaches that integrate quantum metrology and quantum communication techniques have been proposed \cite{HMM:19:PRA, YTZY:20:PRA, SKM:22:PRA, MD:23:AVS, RKKJ:23:SR}. These protocols encompass \textit{secure state transmission} \cite{SKM:22:PRA}, \textit{secure quantum parameter transmission} \cite{XXC:18:QIP,HMM:19:PRA}, and \textit{secure parameter retrieval} \cite{YTZY:20:PRA,RKKJ:23:SR,LCPSP:24:PRA, HSP:25:PRL} of remote systems, 
where the unknown parameter could represent the physical parameter, or might encode variations in optical path length, time delay, or magnetic field intensity. These protocols depend on the random state generation and operations or entanglement between the two parties to provide quantum secure communication.
Recently, entanglement-free quantum remote sensing has been demonstrated in an optical fiber \cite{He:arXiv}.


SQS protocols can leverage entanglement sharing between the source and receiver, followed by subsequent local operations and classical communication (LOCC), to verify security \cite{YTZY:20:PRA,RKKJ:23:SR,LCPSP:24:PRA}. This approach is distinct from other methods, where auxiliary states are added to mask the probe state. In such protocols, the source is not required to generate multiple probe states or apply complex masking operations to secure the communication channel \cite{HMM:19:PRA,SKM:22:PRA, MD:23:AVS}; instead, entanglement alone guarantees probe security, as LOCC operations are sufficient to detect potential eavesdroppers. Typically, Alice prepares the quantum states and encodes the unknown parameter \( \omega \), while Bob—who may represent a device with limited capabilities—is neither required to prepare quantum states nor know the parameter \( \omega \). Bob acts solely as a remote sensor, located in an environment where we would have restricted access to the system of interest, or as an integral node within a sensing network that communicates securely with Alice, the network's central node, responsible for coordinating the quantum communication and controlling the sensing process.

In this paper, we introduce an \( N \)-partite quantum remote sensing protocol capable of achieving HL precision while ensuring information-theoretic security. To further address realistic scenarios that include noise, we extend our protocol by incorporating a quantum optimal control (QOC) method based on local quantum operations. We demonstrate the effectiveness of our QOC-enhanced protocol by analyzing its performance under various noise models using a three-qubit system.

The rest of this paper is organized as follows. Section~\ref{sec:prelim} provides basic background information and key preliminary concepts. Section~\ref{sec:protocol} introduces our general \( N \)-partite quantum sensing protocol for an ideal quantum environment. Section~\ref{sec:noisy_cqrs} describes the application of our protocol to a noisy scenario involving a three-qubit Greenberger–Horne–Zeilinger (GHZ) state, where we combine parameter-dependent evolution and QOC operations while 
monitoring entanglement through the tripartite negativity measure. Finally, we summarize our findings and conclude the paper in section \ref{conclusion}.



\section{Preliminaries} \label{sec:prelim}

In this section, we review the key principles of quantum metrology under noisy conditions and then introduce the fundamental concepts of QOC for enhancing parameter estimation.

\subsection{Quantum Metrology}
In a quantum metrology protocol, an unknown parameter \( \omega \) is estimated by first preparing a suitable probe state and allowing it to evolve under parameter-dependent dynamics. Subsequently, a measurement described by a positive operator-valued measure (POVM) \( \{M_x\} \), satisfying \( \sum_x M_x = \mathbb{I} \), is performed, and the measurement outcomes are post-processed to infer the parameter value.

For a given POVM, the probability of obtaining outcome \( x \) is given by
\[
p(x|\omega)=\mathrm{Tr}[M_x \rho_\omega],
\]
where \( \rho_\omega \) is the state dependent on the parameter \( \omega \). The sensitivity of the outcome statistics with respect to changes in \( \omega \) is characterized by the classical Fisher information (CFI), which can be written as

\begin{equation} \label{eq:CFI}
\begin{aligned}
\mathcal{F}_C(\omega) = \sum_x \frac{1}{\mathrm{Tr}[M_x \rho_\omega]} \left( \mathrm{Tr}\!\Big[M_x \frac{\partial \rho_\omega}{\partial \omega}\Big] \right)^2.
\end{aligned}
\end{equation}

If one performs \(\nu\) independent repetitions of the protocol, the precision of the parameter estimate is bounded by the classical Cramér-Rao bound:

\begin{align}
\Delta \tilde{\omega}_C \geq \frac{1}{\sqrt{\nu \, \mathcal{F}_C(\rho_{\omega})}}.
\end{align}

Optimizing the measurement over all possible POVMs yields the maximum achievable information about \(\omega\). This optimization leads to the concept of the quantum Fisher information (QFI), denoted by \(\mathcal{F}_Q(\rho_\omega)\). Accordingly, the ultimate precision limit for any quantum measurement is given by the quantum Cramér-Rao bound (QCRB) \cite{GLM:11:Nature, TA:14:JPA, DM:14:PRL}:
\begin{equation} \label{eq:QCRB}
\Delta \tilde{\omega}_Q \geq \frac{1}{\sqrt{\nu \, \mathcal{F}_Q(\rho_{\omega})}},
\end{equation}
where \(\rho_{\omega_0}\) is the state at the true value \(\omega_0\). The QFI thus quantifies the maximum amount of information attainable about the parameter \(\omega\) when optimized over all measurements and is defined as 

\begin{align} \label{eq:QFI}
\mathcal{F}_Q(\rho_{\omega}) = \text{Tr}[\rho_{\omega} L_{\omega}^2],
\end{align}
where \( L_{\omega} \) is the \textit{symmetric logarithmic derivative} (SLD), implicitly defined via the relation
\[
\partial_\omega \rho_{\omega} = \frac{1}{2}(L_{\omega} \rho_{\omega} + \rho_{\omega} L_{\omega}).
\]

Explicitly, the SLD can be written in the eigenbasis of \( \rho_{\omega} = \sum_i \lambda_i(\omega) |e_i(\omega)\rangle \langle e_i(\omega)| \) as
\begin{equation}
L_\omega = \sum_{i,j \,:\, \lambda_i + \lambda_j \neq 0} 
\frac{2 \, \langle e_i(\omega) | \dot{\rho}_\omega | e_j(\omega) \rangle}
{\lambda_i(\omega) + \lambda_j(\omega)} 
\, | e_i(\omega) \rangle \langle e_j(\omega) |,
\end{equation}
where \( \dot{\rho}_\omega = \partial_\omega \rho_\omega \) is the derivative of the state with respect to the parameter.

In realistic scenarios, the evolved quantum state \( \rho_{\omega} \) incorporates noise and decoherence, and is obtained via a completely positive trace-preserving map \( \Lambda_\omega(t) \) acting on the initial state \( \rho_0 \):
\begin{equation} \label{eq:NoisyEvolution}
\rho_{\omega} = \Lambda_{\omega}(t)[\rho_0].
\end{equation}
This map can often be decomposed as \( \Lambda_{\omega}(t) = \mathcal{U}_{\omega}(t) \circ \Gamma(t) \), where:
\begin{itemize}
  \item \( \mathcal{U}_{\omega}(t) = e^{-i H_{\omega} t} \, \boldsymbol{\cdot} \, e^{i H_{\omega} t} \) is the unitary superoperator describing the ideal, coherent evolution under the parameter-dependent Hamiltonian \( H_\omega \),
  \item \( \Gamma(t) \) is the decoherence superoperator, capturing the effects of environmental noise.
\end{itemize}

\begin{figure*}[htbp] 
    \centering
    \includegraphics[width=0.85\textwidth]{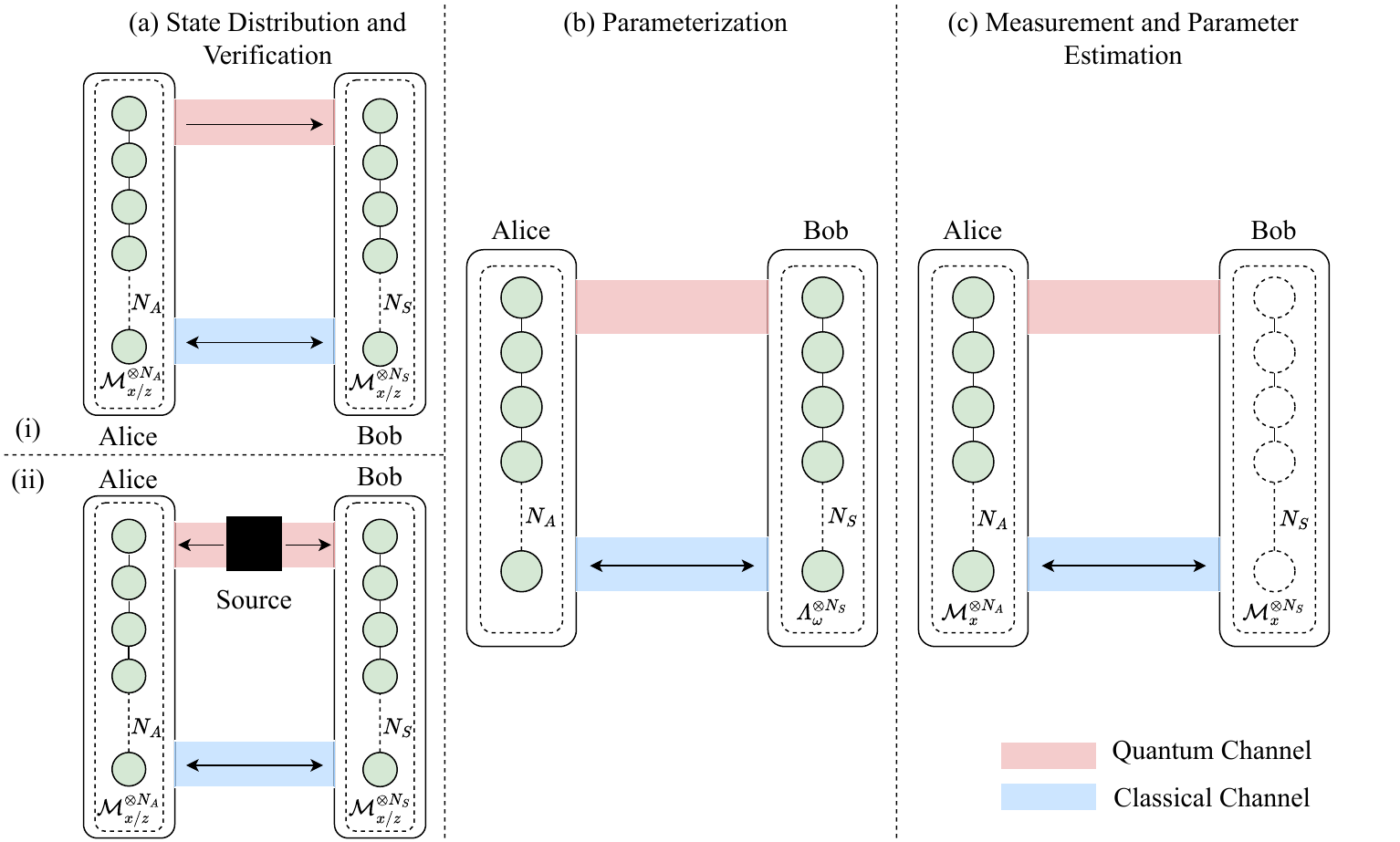}  
    \caption{\raggedright The C-QSRS protocol for $N$ particles. The process has three steps: (a) the state distribution and verification process, which includes either (i) Alice or (ii) an external source; (b) the parameterization process in which Bob encodes the $\omega$-dependent quantum channel, and (c) the measurement and parameter estimation process. Alice and Bob are connected by a quantum and classically authenticated channel, with the arrows representing the presence and direction of the information flow. 
    }
    \label{fig:combined_C-QSRM}
\end{figure*}

\begin{table*}[ht]
\centering
\caption{Summary of noise models for two configurations: (a) when Alice acts as the quantum state source, and (b) when the source is external. Each model specifies a sequence of noise operations affecting the quantum channel.}
\label{tab:noise_models}
\begin{tabular}{p{5.0cm}@{\hspace{1.2cm}}p{2.6cm}p{8cm}}
\toprule
\textbf{Noise Model} & \textbf{Abbreviation} & \textbf{Description} \\
\midrule
\multicolumn{3}{l}{\textbf{(a) Alice as the Source}} \\
\midrule
Parallel Pauli Dephasing                           & PPD        & Noise model where Pauli dephasing occurs in parallel on Bob’s qubits. \\
Generalized Pauli Dephasing                        & GPD        & A more general form of Pauli dephasing applied on Bob’s side. \\
Depolarizing Noise                                 & DP         & Noise model where the state is depolarized, representing a uniform loss of information. \\
Depolarization + Parallel Pauli Dephasing          & DP + PPD   & Imperfect quantum channel with noise applied over the whole state, followed by parallel Pauli dephasing. \\
Depolarization + Generalized Pauli Dephasing       & DP + GPD   & Imperfect quantum channel with noise applied over the whole state, followed by generalized Pauli dephasing. \\
Depolarization + Depolarizing Noise                & DP + DP    & Two consecutive applications of depolarizing noise. \\
\midrule
\multicolumn{3}{l}{\textbf{(b) External Source}} \\
\midrule
Asymmetrical Depolarization + Parallel Pauli Dephasing      & ADP + PPD  & Imperfect quantum channel with noise applied asymmetrically over the state before parallel Pauli dephasing. \\
Asymmetrical Depolarization + Generalized Pauli Dephasing   & ADP + GPD  & Imperfect quantum channel with noise applied asymmetrically over the state before generalized Pauli dephasing. \\
Asymmetrical Depolarization + Depolarizing Noise            & ADP + DP   & Imperfect quantum channel with noise applied asymmetrically over the state before depolarizing noise is applied. \\
\bottomrule
\end{tabular}\label{table:table1}
\end{table*}

\subsection{Quantum Optimal Control}

To steer a quantum system, we supplement its intrinsic Hamiltonian with time-dependent control terms generated by external fields.  Thus, the total Hamiltonian takes the form
\begin{equation}
H(t) = H_{0}(\omega) + H_{c}(t)\,,
\end{equation}
where \(H_{0}\) is the unperturbed Hamiltonian characterized by the parameter \(\omega\), and \(H_{c}(t)\) encodes the influence of our control pulses.

When the system interacts with its environment, its density matrix \(\rho(t)\) evolves according to a Lindblad master equation,
\begin{equation}
\dot{\rho}(t) = \mathcal{L}[\rho(t)]\,,
\end{equation}
with the superoperator \(\mathcal{L}\) accounting for both coherent dynamics and dissipative effects.  Explicitly, one can write
\begin{align}
\dot{\rho}(t) &= -\frac{i}{\hbar}\bigl[H(t),\,\rho(t)\bigr] \nonumber\\
&\quad + \sum_{k} \gamma_k(t)\,\Bigl(L_k\,\rho(t)\,L_k^\dagger - \tfrac{1}{2}\{L_k^\dagger L_k,\rho(t)\}\Bigr)\,,
\end{align}
where the \(L_k\) are the Lindblad operators describing specific decoherence channels, and \(\gamma_k(t)\) are their associated rates, which may vary in time.

In this paper, we adopt a memoryless (Markovian) approximation, which is valid when the environmental correlation times are much shorter than the system’s evolution timescale \cite{MM:22:Quantum}.  We partition the total sensing duration \(T\) into \(m\) intervals of length \(\Delta t\), such that
\begin{equation}
\rho(T) = \Bigl(\exp[\Delta t\,\mathcal{L}_m]\Bigr)\cdots\Bigl(\exp[\Delta t\,\mathcal{L}_1]\Bigr)\,\rho(0)\,,
\end{equation}
with each \(\mathcal{L}_i\) held fixed over its interval but allowed to change between steps.

\section{C-QSRS Protocol}\label{sec:protocol}
This section introduces a controlled-quantum secure remote sensing (C-QSRS) protocol under ideal quantum channel conditions, assuming noiseless state distribution and quantum evolution. The goal of the protocol is to allow an experimenter, Alice, to estimate a parameter $\omega$ associated with a remote system owned by Bob -- a semi-trusted party that can perform quantum operations and measurements but cannot generate quantum states.

A key consideration in the protocol design is the method of state preparation. We analyze two possible scenarios:  

\begin{enumerate}
    \item \textbf{External Source Distribution}. A trusted external source generates and distributes entanglement between Alice and Bob.  
    \item \textbf{Alice as the Source}. Alice locally prepares the entangled state and distributes the required qubits to Bob.  
\end{enumerate}

While both cases lead to identical quantum sensing steps under ideal conditions, their distinctions become relevant when we consider noisy quantum channels. The general structure of the proposed protocol is summarized in Fig.~\ref{fig:combined_C-QSRM}. 
The protocol consists of the following steps: 

\begin{itemize}

\item \textbf{State Preparation. }Alice, or an external source, creates a multiparticle GHZ state: 

\begin{equation}
\ket{\Psi_N} = \frac{1}{\sqrt{2}} \left( \ket{0}^{\otimes N} + \ket{1}^{\otimes N} \right)
\end{equation}

where $N = N_A +N_S$, with $N_S$ representing the number of sensing particles and $N_A$ representing the number of ancillary particles. Alice produces $p = p_s + p_c$ number of $\ket{\Psi_N}$, with $p_s$ being the states for the quantum sensing task and $p_c$ states being the ones for checking the security of the quantum channel. For each $\ket{\Psi_N}$ in $p$, Alice sends $N_S$ particles to Bob, while keeping $N_A$ particles to herself.

\item \textbf{Controlled security check. }Alice randomly chooses $p_c$ states from among the $p$ states to perform the security state. From these states, $p_c/2$ states are chosen to be measured in the $\sigma_x$ basis and the other half are chosen for local $\sigma_z$ measurements.

\vspace{0.2 cm}

For the $\sigma_x$ measurement, we have $M =\sigma_{x_1} \otimes \sigma_{x_2} \otimes ........ \otimes \sigma_{x_N}$, with measurement outcomes $\{m_{x_1}, m_{x_2},.........,m_{x_N}\}$, with $m_{x_i}=\{1,-1\}$. Bob reveals the results of his measurements using the classical channel, which Alice receives and uses her results to perform the parity check:

\begin{equation}
S_N =  \prod_{k=1}^{N} m_{x_k}
\end{equation}

If $S_N = 1$, the quantum channel passes the $\sigma_x-$basis test, and if $S_N=-1$, the test fails. 

\vspace{0.2 cm}

For the $\sigma_z$ basis measurement, we check if the measurement outcomes $\{m_{z_1},m_{z_2},........,m_{z_N}\}$ with $m_{z_i} = \{-1,+1\}$, the results should be equal. If the outcomes $m_{z_i}$ are not equal, the channel fails the $\sigma_z$-basis test. If the quantum channel passes both tests, we can move to the quantum sensing part of the protocol.

\item \textbf{Controlled QM. }For controlled quantum metrology, we have Alice utilizes $p_s$ states to perform the sensing protocol. We assume that the system Hamiltonian is defined by:

\begin{equation}
    H_\omega = \frac{\omega}{2} \sum_{j=1}^{N_S} \sigma_{z}^{(j)}.
\end{equation}

Bob evolves the system under $H_\omega$ for time $t_s$, after which he performs a local $\sigma_x$-basis measurement.

\begin{figure*}[htbp]  
    \centering
    \includegraphics[width=0.85\textwidth]{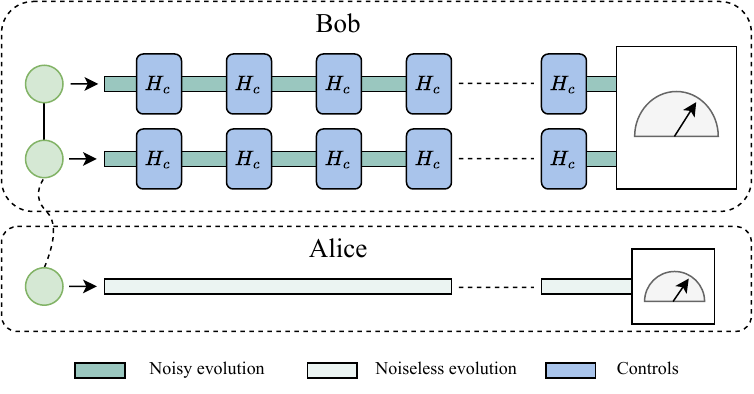}  
    \caption{The QOC process applied locally by Bob, who possesses the noisy channel to be probed. After the evolution, Alice and Bob perform measurements to ascertain the parameter. }
    \label{fig:combined_evolution}
\end{figure*}

\item \textbf{Secure parameter teleportation. 
}The final state after evolution becomes:

\begin{equation}
\small
\ket{\Psi_N}_{\omega} = \frac{1}{\sqrt{2}} 
\left( e^{\frac{i N_S \omega t_s}{2}} \ket{0}^{\otimes N} + e^{\frac{-i N_S \omega t_s}{2}} \ket{1}^{\otimes N} \right).
\end{equation}

Bob is instructed to perform his measurements in the $\sigma_x$-basis, after which the state becomes

\begin{equation}
\small
\ket{\Psi_N}_{\omega} = \frac{1}{\sqrt{2}} 
\left( e^{\frac{i N_S \omega t_s}{2}} \ket{0}^{\otimes N_A} \pm e^{\frac{- i N_S \omega t_s}{2}} \ket{1}^{\otimes N_A} \right),
\end{equation}

which can be used to estimate the parameter.

\item \textbf{Parameter estimation. } Alice performs her $\sigma_x$-basis measurement, and she can use the results to reconstruct the parameter estimate $\tilde{\omega}$. The QCRB becomes:

\begin{align}
(\Delta \tilde{\omega})^2 &\approx 
\frac{1}{p_s} 
\frac{(\Delta \sigma_x)^2_{\text{id}}}
     {\left| \partial_\omega \langle \sigma_x \rangle_{\text{id}} \right|^2} \nonumber \\
&= \frac{1}{p_s} 
\frac{\sin^2(N_S\omega t_s)}
     {\left| N_S t_s \sin(N_S \omega t) \right|^2} =
\frac{1}{p_s N_{S}^2 {t_s}^2},
\end{align}

which approaches the HL. Using this approach, Bob cannot retrieve any information about the parameter.

\item \textbf{Parameter integrity. }We will now check the security of the encoded parameter as visualized by Eve. In this context, we observe that the parameter encoding process only starts once the integrity of the transported GHZ state is verified, and after measurement by Bob, no parameter information remains with him. To verify this, we observe that the available system to Bob after encoding the parameter is $\rho_{\omega}^{AB} = \ket{\psi_N}\bra{\psi_N}_{\omega}$. After tracing out the subsystem possessed by Alice, the state possessed by Bob becomes

\begin{equation}
\text{Tr}_{A} (\rho_{\omega}^{AB}) = \frac{1}{2}\ket{0}\bra{0}^{\otimes N_S} + \frac{1}{2}\ket{1}\bra{1}^{\otimes N_S}
\end{equation}

which eliminates any parameter information that Bob might gain. Another attack that can compromise the integrity of the estimated parameter is an induced bias $\beta$, by Eve, that can hamper the estimation precision as the total parameter value becomes $\omega + \beta$, with $\beta$ being unknown. However, we observe that such tampering can be detected when Pauli-X measurements are performed for the state verification process. 

\end{itemize}


The parameters \( t_s \), \( N_A \), and \( N_S \) can be adjusted to optimize the protocol according to a specific encoding strategy \cite{HMM:16:PRA, HMM:18:PRA, LHYY:23:PRA}. Additionally, in scenarios where an untrusted external source is used for state preparation and entanglement distribution, the verification step at Alice and Bob’s can include checking stabilizer conditions to fully verify the incoming state \cite{PCW:12:PRL, SM:22:PRA}. However, this approach would necessitate a higher value of \( p_c \), thereby reducing the number of resources available for the sensing process and, consequently, the achievable sensitivity when the resources are finite.


\begin{figure*}[htbp]  
    \centering
        \includegraphics[width=1.05\textwidth]{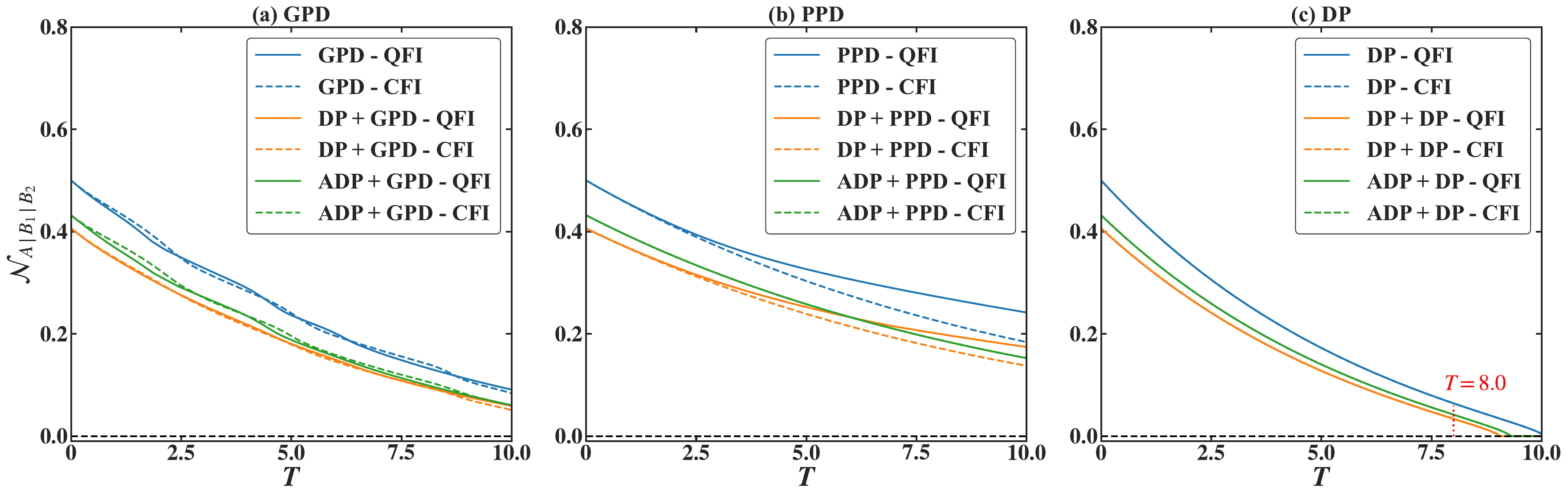}  
    \caption{Comparison of the tripartite negativity ($\mathcal{N}_{A\mid B_1\mid B_2})$ as a function of time ($T$) under (a) GPD noise, (b) PPD noise, and (c) DP noise for the two scenarios considered in this work: noiseless communication with noisy evolution, and noisy communication with noisy evolution. }
    
    \label{fig:combined_graph_neg}
\end{figure*}

\section{Noisy C-QSRS}\label{sec:noisy_cqrs}
In this section, we augment QOC with our C-QSRS protocol for the case of non-ideal dynamics. To keep our analysis both insightful and computationally manageable, we focus on a three-particle system ($N=3$), with $N_A=1$ and $N_S=2$, which results in the initial state:

\begin{align}\label{eq:init_state}
\ket{\Psi}_{A B_1 B_2} = \frac{1}{\sqrt{2}} \left( \ket{000} + \ket{111} \right)_{A B_1 B_2},
\end{align}

In our setup, both Alice and Bob are assumed to have perfect quantum memories. Within our noisy framework, we compare systems that operate without any intervention (uncontrolled) with those that benefit from QOC. In the controlled systems, local QOC operations are used to actively counteract the noise, leading to enhanced overall performance. The cases are discussed in detail in Table ~\ref{table:table1}.

\subsection{Framework}

\subsubsection{Control Hamiltonian}

To counteract the detrimental effects of noise, we employ a time-dependent local quantum optimal control Hamiltonian of the form
\begin{equation}
H_c(t) = \sum_{i=1}^3 u_i(t) \sigma_i^{(2)} + \sum_{j=1}^3 v_j(t) \sigma_j^{(3)},
\end{equation}
where $\{\sigma_0, \sigma_1, \sigma_2, \sigma_3\} = \{I, \sigma_x, \sigma_y, \sigma_z\}$ and $\sigma_i^{(n)}$ denotes the $i$-th Pauli operator acting on the $n$-th qubit. Here, the control acts locally on Bob's two qubits (labeled as the 2nd and 3rd qubits in the system), with $u_i(t)$ and $v_j(t)$ being the time-dependent control amplitudes optimized to maximize the chosen figure of merit. Alice’s qubit remains free of control, with the application of QOC operations only focused on the evolution at Bob's side, as illustrated in Fig.~\ref{fig:combined_evolution}.

We use the QFI and the CFI as the objective functions. For the CFI optimization, the local observable $\sigma_x$ is used. For the cases of generalized and parallel Pauli dephasing, we use the Gradient Ascent Pulse Engineering Algorithm (GRAPE) \cite{storn1997differential,KRKH:05:JMR,LY:17:PRA,ZYYW:22:PRR} as the optimization algorithm.  Differential Evolution is employed to deal with the Depolarization noise \cite{,lovett2013differential,ZYYW:22:PRR}. 

\subsubsection{Tripartite Negativity}

In our analysis, we benchmark the optimal evolution time by quantifying genuine tripartite entanglement using tripartite negativity. For a system of three qubits \(A\), \(B_1\), and \(B_2\), we compute negativities for the bipartitions
\[
A \mid B_1B_2,\quad
AB_1 \mid B_2,\quad
AB_2 \mid B_1.
\]
For any bipartition \(X\mid Y\), the bipartite negativity is defined as \cite{VW:02:PRA}
\[
\mathcal{N}_{X\mid Y} = \frac{\bigl\|\rho_{XY}^{T_X}\bigr\|_1 - 1}{2},
\]
where \(\rho_{XY}^{T_X}\) denotes the partial transpose with respect to subsystem \(X\), and \(\|\cdot\|_1\) is the trace norm. The tripartite negativity is then the geometric mean of these three \cite{SG:08:EPJD}:
\begin{equation}\label{eq:negativity}
\mathcal{N}_{A\mid B_1\mid B_2}
= \sqrt{\mathcal{N}_{A\mid B_1B_2}
        \;\mathcal{N}_{AB_1\mid B_2}
        \;\mathcal{N}_{AB_2\mid B_1}}.
\end{equation}
This measure provides a robust benchmark for the maximum evolution time before entanglement vanishes.

\subsection{Communication Scenarios}
The communication scenarios that we discuss consist of two distinct scenarios: one featuring noiseless communication combined with noisy encoding, and another involving both noisy communication and noisy encoding.

\subsubsection{Noiseless Communication and Noisy Encoding}
In this scenario, we assume that the communication channels exhibit little to no noise. This situation is plausible when Alice and Bob implement effective entanglement purification steps \cite{,RRR:24:Quantum,CLL:07:PRA} prior to the noisy evolution stage. Specifically, the initial state is prepared as a tripartite GHZ state as in \ref{eq:init_state}, with one qubit distributed to Alice and the remaining two sent to Bob. Although each of the three qubits may experience slight noise during transmission, subsequent entanglement purification procedures are assumed to restore their ideal entangled correlations. 

\begin{figure*}[htbp]  
    \centering
    \includegraphics[width=0.95\textwidth]{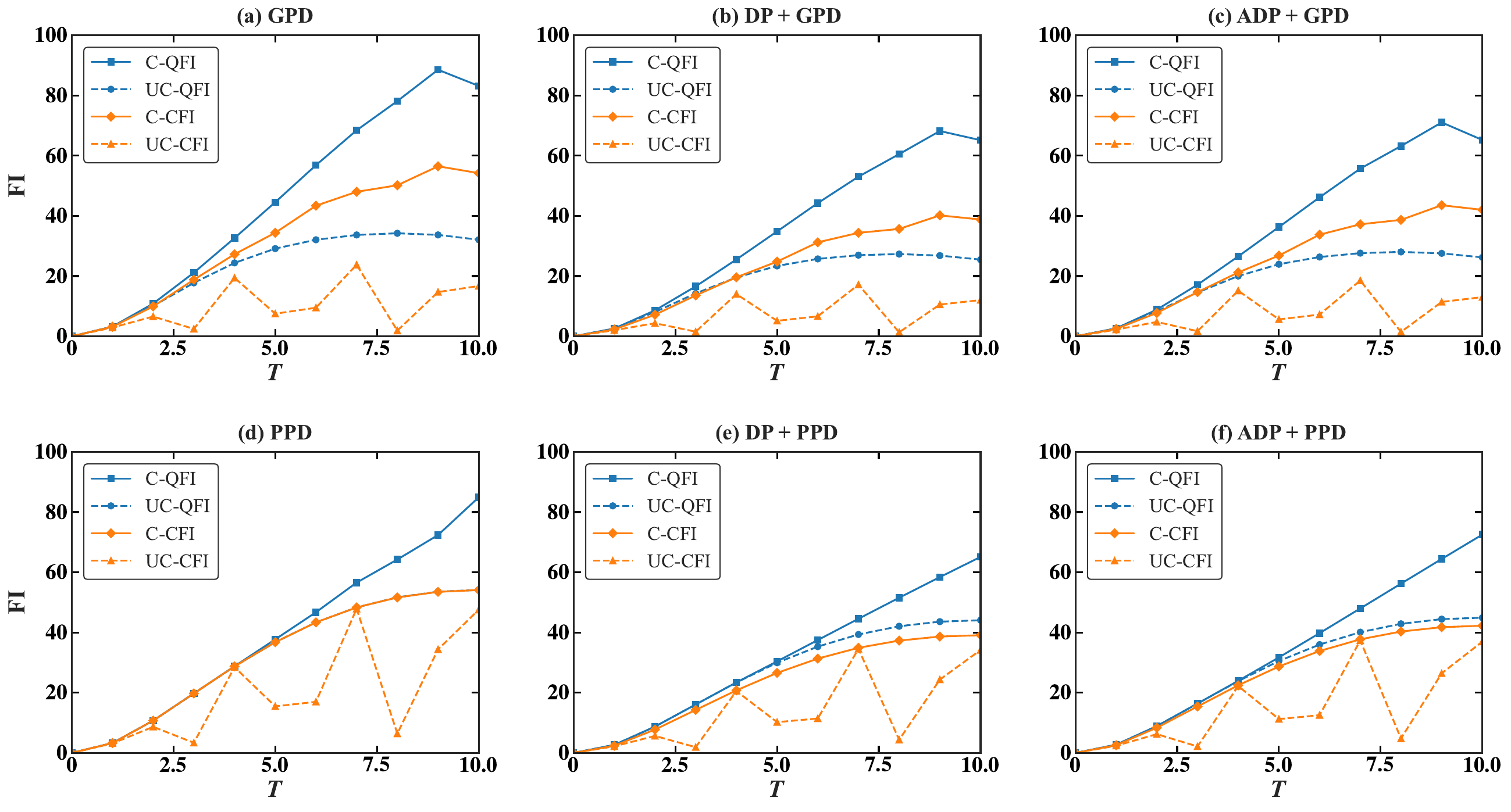}  
    \caption{Comparison of QFI and CFI for the cases of (i) noiseless communication and noisy evolution, (ii) noisy communication and noisy evolution.  }
    \label{fig:combined_graph}
\end{figure*}

\subsubsection{Noisy Communication and Noisy Evolution}
In this scenario, we have two cases based on whether Alice is the source or the source is external. 
\vspace{0.2cm}
\paragraph{External Source.}
In this scenario, the communication channel introduces uniform noise to each qubit via a depolarizing channel. The channel is characterized by the map
\[
\mathcal{E}(\rho) = (1 - \lambda) \rho + \lambda \frac{\mathcal{I}}{d},
\]
where $\rho=\ket{\Psi}\bra{\Psi}$, with $\ket{\Psi}$ from Eq.~\eqref{eq:init_state},  \( \lambda \) quantifies the noise strength and \( d \) is the dimension of the Hilbert space. Under the action of this depolarizing channel, the initial GHZ state evolves into a Werner state which is subsequently incorporated into the master equation to analyze the system's noisy evolution.

\vspace{0.2cm}
\paragraph{Alice as a Source.}

In the scenario where Alice serves as the source, Alice retains a portion of the quantum state while the remaining qubits are transmitted through the noisy channel. Here, the depolarization channel acts asymmetrically, affecting only the transmitted parts of the state. This asymmetrical application is modeled by

\begin{align*}
\mathcal{E}(\rho) = \sum_{i,j} \left( K_i \otimes K_j \otimes I_2 \right) \, \rho \, \left( K_i^\dagger \otimes K_j^\dagger \otimes I_2 \right),
\end{align*}
with the Kraus operators \( K_i \) defined as:
\begin{align*}
K_0 &= \sqrt{1 - \frac{3 \Gamma}{4}}\,I,\quad & 
K_1 &= \sqrt{\frac{\Gamma}{4}}\,X,\\[6pt]
K_2 &= \sqrt{\frac{\Gamma}{4}}\,Y,\quad & 
K_3 &= \sqrt{\frac{\Gamma}{4}}\,Z.
\end{align*}
where \( \Gamma \) represents the depolarization strength. The action of these operators on the GHZ state results in a density matrix with modified elements:

\begin{equation}\label{eq:asym_rho_elements}
\begin{aligned}
\rho_{1,1} &= \rho_{8,8} &&= \frac{1}{16}\Bigl(|(4-3\Gamma)\Gamma| + (5\Gamma-12)\Gamma + 8\Bigr),\\
\rho_{2,2} &= \rho_{7,7} &&= \frac{\Gamma^2}{8},\\
\rho_{3,3} &= \rho_{6,6} &&= \frac{1}{16}\Bigl(|(4-3\Gamma)\Gamma| + \Gamma^2\Bigr),\\
\rho_{4,4} &= \rho_{5,5} &&= \frac{1}{16}\Bigl(|(4-3\Gamma)\Gamma| + \Gamma^2\Bigr),\\
\rho_{1,8} &= \rho_{8,1} &&= \frac{1}{16} \Bigl(-| 4-3 \Gamma |  | \Gamma | +\Gamma  (5 \Gamma -12)+8\Bigr)
\end{aligned}
\end{equation}

This asymmetrically depolarized state, reflecting the scenario in which only the communicated portion is exposed to noise, is subsequently employed in the QOC application for the case with Alice as the source, with $\Gamma = 0.06$.

\subsection{Noise Evolution}
Now we introduce three representative noise models that characterize the decoherence affecting Bob’s qubits during their evolution. We analyze the effectiveness of simultaneous local QOC operations in mitigating these effects. Our analysis spans a range of evolution times, \(T = 1, 2, \dots, T_{\text{f}}\), where \(T_{\text{f}}\) is defined as the evolution time before the entanglement is completely lost (i.e., when \(\mathcal{N}_{A\mid B_1\mid B_2} = 0\)).

\subsubsection{Generalized Pauli Dephasing Noise}
Generalized Pauli dephasing (GPD) noise in quantum metrology involves phase fluctuations affecting the quantum state along multiple axes. In this model, dephasing is not restricted to a single direction but is represented by a combination of the Pauli matrices \( \sigma_x \), \( \sigma_y \), and \( \sigma_z \). 

The dynamics of a system subject to GPD noise are governed by the following master equation:
\[
\frac{d\rho}{dt} = -i[H, \rho] + \sum_{k=1}^3 \left( L_k \rho L_k^\dagger - \frac{1}{2} \{ L_k^\dagger L_k, \rho \} \right),
\]
where the Lindblad operators \( L_k \) capture the dephasing effects. In our model, these operators are defined as:
\[
L_k = \sqrt{\gamma_{gp}} \left( \sigma_x \sin(\theta) \cos(\phi) + \sigma_y \sin(\theta) \sin(\phi) + \sigma_z \cos(\theta) \right),
\]

with \( \theta = \pi/4 \), \( \phi = 0 \), and $\gamma_{gp}$, which we set at \(\gamma_{gp} = 0.05\) representing the dephasing rate. 

We first examine the tripartite negativity in Figure~\ref{fig:combined_graph_neg}(a), which compares three scenarios under the GPD noise model: (a) pure GPD, (b) depolarizing noise followed by GPD (DP + GPD), and (c) asymmetrical depolarizing noise followed by GPD (ADP + GPD). Notably, the QOC operations do not improve the tripartite negativity—the controlled cases closely follow the trends of their uncontrolled counterparts.

Next, Figures~\ref{fig:combined_graph}(a-c) present the Fisher information (FI) performance. The results clearly show that QOC significantly enhances both the QFI and the CFI, with the controlled metrics (C-QFI and C-CFI) consistently outperforming the uncontrolled ones (UC-QFI and UC-CFI). In particular, the pure GPD scenario exhibits the highest performance, whereas the introduction of depolarizing noise degrades the precision. Interestingly, C-CFI even surpasses UC-QFI, indicating that Alice’s parameter estimate—based on her and Bob’s local measurement data —can outperform the best uncontrolled strategy. 

\vspace{0.2cm}

\subsubsection{Parallel Pauli Dephasing Noise}
Parallel Pauli dephasing (PPD) noise acts along the same direction as the parameter-encoding Hamiltonian—i.e  \( z \)-axis of the Bloch sphere. This noise mechanism leads a decay of quantum coherence, characterized by a reduction of the off-diagonal elements of the density matrix, while leaving the diagonal elements representing the state populations unaffected, and is especially hard to mitigate \cite{SSK:17:Quantum,ZZP:18:Nature}. 

The dynamics of the system under parallel dephasing noise are modeled by the master equation:
\begin{equation}
\frac{d\rho}{dt} = -i[H, \rho] + \gamma_z \left( \sigma_z \rho \sigma_z^\dagger - \frac{1}{2} \{ \sigma_z^\dagger \sigma_z, \rho \} \right),
\end{equation}
where \( \gamma_z\), which we set at \( \gamma_z= 0.025 \), quantifies the strength of the noise in the communication channel, and \(\sigma_z\) is the operator corresponding to dephasing along the \(z\)-axis.

\begin{figure*}[htbp]  
    \centering
    \includegraphics[width=0.95\textwidth]{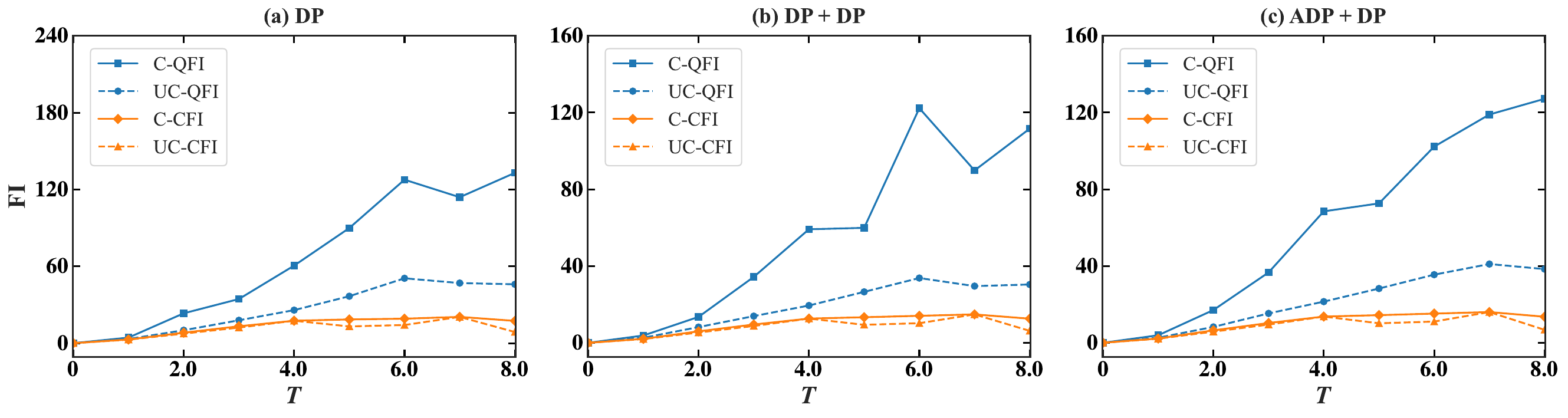}  
\caption{The performance of QFI and CFI under depolarizing noise represented by the Fisher information (FI) versus total evolution time \(T\) for three channel configurations: (a) noiseless channel, (b) external source (DP), and (c) Alice as source (ADP +DP).}

    \label{fig:combined_graph_dep}
\end{figure*}

Figure~\ref{fig:combined_graph_neg}(b) shows the tripartite negativity for three cases under parallel dephasing noise: PPD, PPD with an additional depolarizing channel (DP + PPD), and asymmetrical depolarizing noise followed by PPD (ADP + PPD). In the PPD and DP + PPD cases, the application of QOC yields a modest improvement, reducing the rate at which the tripartite negativity declines compared to the uncontrolled schemes.

Figure~\ref{fig:combined_graph}(d-f) illustrates the evolution of the QFI and CFI for the same noise models. In every scenario, QOC substantially enhances performance, as evidenced by the controlled metrics (C-QFI and C-CFI) consistently surpassing their uncontrolled counterparts (UC-QFI and UC-CFI). While the PPD case maintains relatively robust performance, the introduction of depolarizing noise in the quantum channel (DP + PPD) results in a noticeable reduction in both QFI and CFI. The ADP + PPD scenario exhibits similar degradation, with asymmetrical depolarization compromising the state and leading to an almost equivalent performance.

\subsubsection{Depolarizing Noise}
Depolarizing noise randomizes the state of a quantum system by mixing it with the maximally mixed state, thereby degrading coherence and entanglement. This degradation leads to a reduction in the precision of parameter estimation in quantum metrology, making the understanding and mitigation of depolarizing noise essential for enhancing the performance of quantum sensors.

The dynamics of a quantum system under depolarizing noise is described by the Lindblad master equation \cite{MCM:13:PRA}:
\begin{equation}
\frac{d\rho}{dt} = -i[H, \rho] + \gamma_d \sum_{k=1}^3 \left( L_k \rho L_k^\dagger - \frac{1}{2} \{ L_k^\dagger L_k, \rho \} \right),
\end{equation}
where the Lindblad operators are given by:
\begin{align*}
L_1 &= \sigma_x, \quad L_2 = \sigma_y, \quad L_3 = \sigma_z,
\end{align*}
and \(\gamma_d\) represents the depolarizing noise rate and we set it as \(\gamma_d=0.02\).

We first observe the tripartite negativity in Figure~\ref{fig:combined_graph_neg}(c), which compares three cases under depolarizing noise: (a) pure depolarizing noise (DP), (b) two consecutive depolarizing channels (DP + DP), and (c) asymmetrical depolarization followed by depolarizing noise (ADP + DP). Notably, QOC operations provide no advantage in enhancing entanglement, and the tripartite negativity reaches zero for almost all cases before or at \( T = 10 \). Consequently, to run the protocol while entanglement is sustained, we select \( T_f = 8.0 \) as the evolution time.

Figure~\ref{fig:combined_graph_dep}(a-c) displays three graphs corresponding to distinct depolarizing noise models. In the pure DP scenario, local QOC markedly enhances both the QFI and the CFI, with the C-QFI substantially exceeding the UC-QFI, and the C-CFI showing moderate improvement. When a second depolarizing channel is introduced (DP + DP), the overall performance is further degraded; however, QOC still yields a noticeable enhancement in QFI. The ADP + DP case exhibits a similar trend, demonstrating that even under compounded depolarizing conditions, the application of QOC consistently improves the estimation performance.

\section{Conclusion}\label{conclusion}
In this paper, we have introduced an \( N \)-partite C-QSRS protocol that achieves HL precision while ensuring unconditional security through the use of shared multipartite entanglement between two remote parties. We demonstrated how a basic version of the protocol can be enhanced via QOC techniques to increase the attainable Fisher information, even under the degrading influence of environmental decoherence.

In the noisy C-QSRS scenario, we assumed local measurements in the \( \sigma_x \) basis and optimized the CFI accordingly. However, the framework remains adaptable: other measurement observables and LOCC strategies may be employed to further optimize information extraction. The resulting C-CFI is always upper bounded by the C-QFI, which we evaluated for various noise models.

Our protocol also lays the groundwork for practical quantum remote sensing implementations, particularly as entanglement generation technologies continue to advance. Our protocol is designed with flexibility in mind, supporting arbitrary entangled particle number \( N \), tunable encoding schemes, and general ancilla dimension \( N_A \), which may be controlled. These degrees of freedom allow the protocol to be tailored to specific applications. For instance, higher-dimensional ancilla systems or more extensive multipartite entanglement can potentially enhance robustness against noise \cite{NBCA:16:PRA,WWZB:18:PRA,KGAD:23:PRL}. We believe that this work represents a meaningful step toward secure and high-precision quantum sensing in realistic, noisy environments.

Future work could explore generalizing the application of QOC to larger quantum systems shared between distant parties, though doing so introduces substantial computational complexity due to the exponential growth of the Hilbert space \cite{KBCD:22:EPJ,LJSK:24:JPC}. Additionally, an important extension would be to investigate the performance of C-QSRS under non-Markovian noise and develop strategies to mitigate its impact.

\section*{Disclosures}
The authors declare that they have no known competing financial interests.

\section*{Data availability}
\vspace{-1em}
The code and datasets generated and analyzed during this study are available from the corresponding author upon reasonable request.

\bibliographystyle{apsrev4-2} 

\bibliography{manuscript}        

\end{document}